CHAPMAN UNIVERSITY | INSTITUTE FOR QUANTUM STUDIES

**REGULAR PAPER**

Check for updates

# Quantizing Galilean spacetime: a reconstruction of Maxwell's equations in empty space

**Ulf Klein**

**Abstract**  As was recently shown, non-relativistic quantum theory can be derived by means of a projection method from a continuum of classical solutions for (massive) particles. In this paper, we show that Maxwell's equations in empty space can be derived using the same method. In this case, the starting point is a continuum of solutions of equations of motion for massless particles describing the structure of Galilean space-time. As a result of the projection, the space-time structure itself is changed by the appearance of a new fundamental constant $c$ with the dimension of a velocity. This maximum velocity $c$, derived here for massless particles, is analogous to the accuracy limit $\hbar$ derived earlier for massive particles. The projection method can thus be interpreted as a generalized quantization. We suspect that all fundamental fields can be traced back to continuous sets of particle trajectories, and that in this sense, the particle concept is more fundamental than the field concept.

## 1 Introduction

> "…one is struck by the dualism which lies in the fact that the material point in Newton's sense and the field as continuum are used as elementary concepts side by side"
> Albert Einstein [1]

The disturbing dualism between the incompatible basic concepts *particle* and *field* has accompanied modern physics from the very beginning. If one wants to eliminate the particle concept in favor of the field concept, then one needs nonlinear field equations of unknown origin. The reverse way, namely the elimination of the field concept in favor of the particle concept, is much easier to accomplish, at least in a formal way. This elimination is actually an innocent looking standard technique of fluid mechanics known as *Lagrangian to Eulerian transition* [2,3]. It is based on the existence of a continuum of solutions of first order ordinary differential equations which *fill* and thus

U. Klein (✉)
Institute for Theoretical Physics, University of Linz, Altenberger Strasse 69, 4040 Linz, Austria
e-mail: ulf.klein@jku.at

generate the space under consideration [4]; let us note for clarity that the Eulerian formulation is nothing but the standard formulation of physical fields.

A popular version of this deep particle-field duality is the particle-wave duality of quantum theory (QT). Einstein rejected this concept and considered QT to be a theory which is complete only with respect to statistical ensembles and not with respect to individual particles. From the point of view of this ensemble interpretation [5], QT must be understood in terms of its relation to classical statistical ensembles. The general program of a "quantization", or better reconstruction of QT, then consists in the transformation of the basic equations of the theory of classical statistical ensembles into the basic equations and structures of QT. The first step of this transformation must be the Lagrangian to Eulerian transition.

This program has been realized in a series of papers of the present author which will be referred to as I [6], II [3], III [7], and IV [8]. Reconstructing QT along these lines is very simple from a conceptual point of view. It requires, besides the transition to the Eulerian formulation, only two essential steps, namely a projection from phase space to configuration space and a linearization or randomization. Both steps can be motivated physically. A theory containing these two steps, regardless of the order, is referred to as "Hamilton–Liouville–Lie–Kolmogorov theory" (HLLK). This acronym is also used in this paper. As shown in I–IV, the success of the HLLK is remarkable—not only Schrödinger's equation but essentially all characteristic features of (single-particle) QT, like non-commuting observables, Born's rule, and even spin, have been derived.

The program underlying the HLLK has not yet been fully carried out. It is obviously incomplete with respect to the description of several particles and with respect to the extension to Minkowskian spacetime. One wonders, however, if this program cannot be completed in an even more fundamental sense. So far, we have studied only particles of non-zero mass. We have found, in IV, that such particles must be fermions with spin one-half. The second class of elementary particles of fundamental importance in nature are bosons, which have integer spin and are responsible for mediating interactions. The most important structureless particle of this class is the massless photon, which has spin 1, and mediates the electromagnetic interaction. The question arises whether the program of the HLLK can—in analogy to the situation with massive particles—also be used to derive the "quantum mechanical field equations of the photon", i.e., Maxwell's equations. In this work, which is the fifth in this series of papers, we show that this is indeed possible.

The possibility of reconstructing a fundamental *classical* field equation with the help of HLLK represents a surprising extension of the range of validity of this method. This generalization, which concerns fundamental concepts of physics, is discussed in Sect. 2. Here, the interested reader will also find an overview of HLLK as applied to QT so far.

The ordinary QT of massive particles, derived in I to IV, describes the behavior of particles under a variety of external conditions (forces), each corresponding to a particular functional form of the Hamiltonian function. In contrast, in Maxwell's equations (in empty space), there is no quantity whose functional form could be varied; the form of the equations is fixed once and for all. In other words, Maxwell's equations describe only a *single* system. If one asks, what this single system might be, then the only possible (reasonable) answer seems to be: the empty space.

As a consequence, we need, as a starting point of the HLLK program, equations of motion for massless "particles" describing the structure of empty space. Such a differential equation, in which no inertial mass occurs, can be derived by means of a transformation from an inertial system to an arbitrarily moving coordinate system (see Sects. 3 and 4). The result agrees with the equation of motion derived by Holland from Maxwell's equations [9]. In Holland's paper, it is also mentioned that one can, in turn, derive the time-dependent part of Maxwell's equations from these equations of motion, using the standard quantization rules. Since Maxwell's equations are no longer Galilei-invariant, this means that, in the course of this standard quantization, the Galilei invariance is *broken*. In fact, the quantization process introduces, according to Holland, a new fundamental constant $c$ and this implies a fundamental change from Galilean spacetime to Minkowskian spacetime [9]. This is astonishing because physical constructions usually take place *within* a given spacetime. Holland's remarks provided a strong motivation for the author to try to realize the HLLK process for massless particles—despite this mismatch of symmetries which was irritating at first sight.

In this paper, we follow the version of HLLK reported in I and II. The space generated in Sect. 5 by means of the Lagrangian to Eulerian transition is the cotangent bundle of the configuration space $\mathbb{R}^3 \times SO(3)$. We find that we have to exclude a certain subset of all possible trajectories from this process. This implies an additional condition, which leads later, in Sect. 7.3, to the transversality conditions of the electromagnetic fields. It is this additional condition that allows for the first time a derivation of the complete set of Maxwell's equations in empty space.

In Sect. 6, the projection onto the configuration space $\mathbb{R}^3 \times SO(3)$ is performed. This part represents the core of the HLLK quantization. The projection creates a new fundamental constant $c$ with the dimension of a velocity. Due to the additional rotational degrees of freedom, a further projection onto the Euclidean configuration space $\mathbb{R}^3$ is required, which is carried out in Sect. 7. The physical meaning of the system of equations derived in this way is discussed in Sect. 8.

The last Sect. 9 contains some reflections on the concepts of particles, fields, and quantization, as suggested by the success of the HLLK. We propose a new principle for describing the relation between a physical theory and a related "better" theory. This principle may be summarized as follows: "Get rid of idealizing assumptions"

## 2 Structure of the HLLK framework

In this section, we provide for the convenience of the reader a brief overview of the structure of the HLLK, as applied to QT and presented in papers I–IV. We then describe the generalization of the theory on which this paper is based.

Before we start, we should mention some other works that also derive QT using different methods but similar probabilistic assumptions. Numerous such constructions have been published since the discovery of QT, but we will limit ourselves here to some of the more recent works. Building upon the seminal work of Koopman and von Neumann, Bondar et al. [10], Arsiwalla et al. [11], and Chester et al. [12] formulate mappings from classical probabilistic mechanics to QT with slightly different quantization schemes compared to HLLK. Furthermore, based on "indivisible" dynamical systems, Barandes [13,14] has recently proposed a stochastic-quantum correspondence. In [15], Wetterich also discusses a similar idea based on probabilistic cellular automaton. The above approaches are all independent of HLLK, yet convey a similar overarching message as HLLK.

### 2.1 The HLLK framework as applied to quantum theory

The starting point for the reconstruction of QT in the framework of HLLK is a statistical ensemble of classical particle trajectories. This ensemble is mathematically defined as a continuum of solutions of the canonical equations,

$$\dot{q}_k = \frac{\partial H(q, p)}{\partial p_k}, \quad \dot{p}_k = -\frac{\partial H(q, p)}{\partial q_k}, \tag{1}$$

which fills the entire 6-dimensional phase space. We consider a single (classical) particle without interaction, with coordinates $q_k$, conjugate momenta $p_k$, and Hamilton's function $H(q, p)$.

We have thus defined the mathematical structure of the theory that we have chosen as our starting point. The final result of our reconstruction should be the one-particle QT, which in the simplest case, which we consider here first, is given by the Schrödinger equation—a linear evolution equation in three-dimensional configuration space for one complex or two real variables. If we compare the two mathematical structures, the following steps required to reconstruct the QT are almost automatic.

(1) First, the continuous set of particle trajectories must be converted into the usual field-theoretical form. This process is known from fluid mechanics and is called "Lagrangian to Eulerian transition". Field variables are introduced whose temporal and spatial distribution in phase space is determined by the solutions of the equations of motion. The partial differential equations which these field variables obey are best suited for the reconstruction of QT in that they reproduce the behavior of the particles more accurately than any other type of field equation.

(2) The number of field variables to be introduced must correspond to the number of real components of the wave function to be reconstructed. In our case, therefore, two field variables must be introduced. One of these two variables will in any case be a quantity $\rho(q, p, t)$, which we assign the physical meaning of a probability density—in the sense of a quantum mechanical interpretation. It obeys the Liouville equation

$$\frac{\partial \rho}{\partial t} + \frac{\partial \rho}{\partial q_k} \frac{\partial H}{\partial p_k} - \frac{\partial \rho}{\partial p_k} \frac{\partial H}{\partial q_k} = 0. \tag{2}$$

We introduce the classical action in phase space $S(q, p, t)$ as the second field variable. It obeys the equation

$$\frac{\partial S}{\partial t} + \frac{\partial S}{\partial q_k} \frac{\partial H}{\partial p_k} - \frac{\partial S}{\partial p_k} \frac{\partial H}{\partial q_k} = \bar{L}, \tag{3}$$

where $\bar{L}$ is the Lagrangian defined by $\bar{L} == p_k (\partial H / \partial p_k) - H$. This choice is based on the close relationship between the classical action in configuration space and the phase of the quantum mechanical wave function; if (3) is projected onto the configuration space by replacing the variable $p$ by the gradient of $S$, the Hamilton–Jacobi equation is obtained.

(3) In QT, two coupled equations for the probability density and the phase can be derived from the well-known polar representation of the wave function (Madelung decomposition). Conversely, in the context of our reconstruction, we try to obtain with the help of the same decomposition

$$\psi = \sqrt{\rho}\, e^{\frac{i}{\hbar} S}, \tag{4}$$

a single equation for the complex phase space variable $\psi(q, p, t)$. Indeed, we obtain the equation

$$\left[ \frac{\hbar}{\iota} \frac{\partial}{\partial t} - \frac{\hbar}{\iota} \frac{\partial H}{\partial q_k} \frac{\partial}{\partial p_k} + \frac{\partial H}{\partial p_k} \left( \frac{\hbar}{\iota} \frac{\partial}{\partial q_k} - p_k \right) + H \right] \psi = 0. \tag{5}$$

This is still a purely classical statistical theory. In particular, no result of this theory depends on the parameter $\hbar$. It is "only" the mathematical form of this theory that has a similarity with QT; of course, it is this similarity that makes the transition to QT much easier.

(4) Equation (5) is an evolution equation in phase space, but it already fulfills the important criterion of linearity. The final step is to convert (5) into an evolution equation in configuration space. This is formally very easy to do using the "quantization rules"

$$\frac{\partial}{\partial p_k} = 0, \qquad p_k = \frac{\hbar}{\iota} \frac{\partial}{\partial q_k}. \tag{6}$$

The application of these rules transforms (5) into the Schrödinger equation

$$\left[ \frac{\hbar}{\iota} \frac{\partial}{\partial t} + H \left( q, \frac{\hbar}{\iota} \frac{\partial}{\partial q_k} \right) \right] \psi(q, t) = 0, \tag{7}$$

This projection generates far-reaching changes in the theory: The concept of particle orbits becomes obsolete, and the formal parameter $\hbar$ becomes a new natural constant that can no longer be eliminated from the theory.

In this simplest version of the HLLK, we have treated the non-relativistic theory of massive particles without spin. This simplest version can be extended or modified with regard to all three properties. However, two basic conditions must be taken into account in all versions: (1) there is a projection from phase space to configuration space, (2) the resulting equations in configuration space must be linear.

The simplicity of the above version of the HLLK is based on the fact that the evolution equation in phase space is already linear. Thus, in this version, linearization is performed first, as in I, II. In order to take the phenomenon of spin into account, the order must be reversed, i.e., the projection must be performed first, as in III, IV. The projection is performed by replacing the variable $p$ in the canonical equations with a momentum field $M$, $p_k \rightarrow M_k(q, t)$. The canonical equations lead to the nonlinear equation of motion

$$\frac{\partial M_i(q, t)}{\partial t} + \left[ \frac{\partial M_i(q, t)}{\partial q_l} - \frac{\partial M_l(q, t)}{\partial q_i} \right] v_l(q, t) = -\frac{\partial}{\partial q_i} h(q, t), \tag{8}$$

and to an equation for the position coordinate which will not be written down. The fields $h(q, t)$, $v(q, t)$ are defined by $h(q, t) = H(q, M(q, t))$, $v_k(q, t) = V_k(q, M(q, t))$ and the velocity field $V_k(q, p)$ is given by the derivative of $H(q, p)$ with respect to $p_k$. A decomposition of $M$ with respect to Clebsch potentials $P$, $Q$ is then carried out,

$$M_k(q, t) = \frac{\partial S(q, t)}{\partial q_k} + P(q, t) \frac{\partial Q(q, t)}{\partial q_k}, \tag{9}$$

This decomposition doubles the number of real components of the wave function; the two additional components $P$, $Q$ describe the rotational part of the momentum field which is responsible for spin. This then leads to the Pauli–Schrödinger equation.

In the context of HLLK, the spin is thus a collective (rotational) effect of the probabilistic ensemble—similar collective interpretations have been proposed before [16]. The wave function of a spin 1/2 ensemble can be constructed with the help of four (independent) real fields, a probability density $\rho$ and the three components $S$, $P$, $Q$ of the momentum field. The special value 1/2 for the spin is therefore a logical consequence of the three-dimensionality of space. A more detailed discussion of the spin phenomenon will be given in a subsequent paper as part of the derivation of the Dirac equation.

## 2.2 The general HLLK framework

This paper shows that the concept of HLLK also makes sense in the case of massless particles. In this case, the object to be quantized is nothing other than empty space. The quantization leads to the free Maxwell equations. The method used is formally similar to that described above, but additional rotational degrees of freedom must be taken into account (in this case the spin arises by a completely different mechanism, namely by projection onto an irreducible representation space). Since we derive a classical equation in this paper, the density $\rho$ cannot be a probability density, but must be interpreted as a real particle density. This freedom in the interpretation of $\rho$ does not only exist in Maxwell's equations, but also in Schrödinger's equation; it opens up the possibility of field quantization.

In I–IV, HLLK is exclusively a method for reconstructing QT, which is associated with the appearance of a new natural constant $\hbar$. The fact that the classical Maxwell equations, containing the constant of nature $c$, can be derived with the same "quantization method" requires a re-evaluation of HLLK. Moreover, this fact calls into question fundamental categories of our physical notions.

There are at least two ways of interpreting this fact. On the one hand, we can retain the concept of quantization and generalize it in such a way that we identify the process of quantization with the process of HLLK. Then all three fields, which are named after Schrödinger, Dirac, and Maxwell, are quantized fields. The special role of $\hbar$ as a quantization feature is omitted; the occurrence of $c$ can just as well be understood as a quantization feature. This semantic definition is supported by the fact that all these fields have equal rights in relation to the second quantization. It is reasonable to assume that all fundamental fields in nature are quantized fields (in this sense). The second possibility is to interpret HLLK not as a quantization process, but as a general method for constructing fundamental physical fields. If one opts for this definition, one can maintain the difference between quantized and classical fields. The remarkable fact that the three fields mentioned can all be derived using the same general method does not, of course, depend on any semantic definition.

The structure of the general HLLK method can be defined as follows:

(1) Find field equations in phase space for *physically meaningful* particle ensembles,
(2) Project these field equations onto the configuration space,
(3) Linearize the field equations in configuration space,

where the order of steps (2) and (3) can be changed. The general structure defined in this way leaves many details of the mathematical implementation open. This concerns the dimension and topology of the considered spaces and the way in which the projection and linearization are carried out. It also concerns the way in which the new natural constants are introduced. In contrast to the usual construction of field equations in the Lagrange formalism and with the help of symmetry considerations, the HLLK is extremely *selective*. For particles with non-zero mass, there are essentially only two physically meaningful particle ensembles that automatically result from the canonical equations in Galilei and Minkowski space. The resulting field equations are the Schrödinger equation (more precisely: the Pauli-Schrödinger equation) and the Dirac equation. The ensemble of massless particles describing Galilean space also appears to be unambiguous. In this case, the HLLK method leads to the free Maxwell equations, as shown in the following sections.

**Fig. 1** This is a generic trajectory for initial values $\mathbf{v}_0$ neither parallel nor perpendicular to $\mathbf{n}$

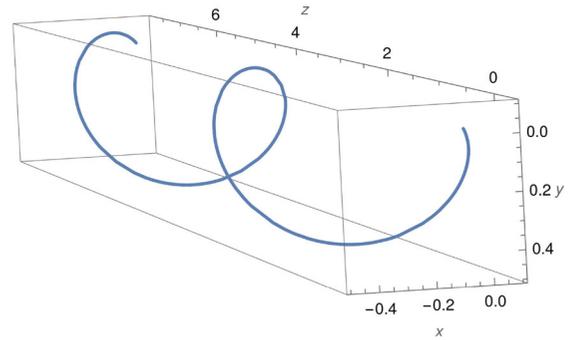

## 3 Equations of motion for Galilean particles

Galilean spacetime is given by $\mathbb{E}^3 \times \mathbb{R}_t$, where $\mathbb{E}^3$ is the Euclidean space equipped with an Euclidean distance function. A complete description of Galilean spacetime is given by its symmetry group, the 10-parameter Galilei group. Those elements of this group that are continuously connected with the identity can be expressed, in a Cartesian coordinate system, in the form

$$(\mathbf{x}, t) \mapsto (R\mathbf{x} + \mathbf{g}t + \mathbf{b}, \, t + s). \tag{10}$$

Here $R$ is a $3 \times 3$ orthogonal matrix with determinant 1, $\mathbf{x}$, $\mathbf{g}$, $\mathbf{b}$ are three-vectors, and $s$ is a scalar parameter. The Galilei group, as Eq. (10) shows, has rotations, Galilei boosts, and space and time translations as subgroups.

Our task is to find equations of motion whose solutions are *as similar as possible* to Eq. (10). We call the hypothetical particles that satisfy these equations *Galilean particles*. We introduce this name in order to emphasize the analogy with both the previous theory of particles with non-zero mass and also with the photons to be derived later. The above definition of the equations of motion is admittedly rather fuzzy; however, we will show that important physical results can be derived with their help.

Let us first consider Newton's equation of motion for a free particle $\ddot{\mathbf{x}} = 0$ and its general solution $\mathbf{x}(t) = \mathbf{v}_0 t + \mathbf{x}_0$. This solution already describes, in its dependence on the initial conditions, two of the above subgroups namely the Galilei boosts and the spatial translations. To account for the still missing rotations, one can tentatively use the defining equation $\dot{\mathbf{x}} = \boldsymbol{\omega} \times \mathbf{x}$ of the angular velocity $\boldsymbol{\omega}$. To recover the Galilei boosts and spatial translations, it is only necessary, to perform a derivation of this equation with respect to time. Considering only angular velocities constant in time (clearly, time-dependent angular velocities are not compatible with our similarity requirement; this would lead to a dramatic increase in the number of degrees of freedom) one obtains the equations

$$\ddot{\mathbf{x}} = \boldsymbol{\omega} \times \dot{\mathbf{x}}, \quad \dot{\boldsymbol{\omega}} = 0, \tag{11}$$

which agree with the equations derived by Holland [9]. These relations have a simple physical interpretation. Let us consider two coordinate systems, an inertial frame $K$, with basis vectors $\mathbf{e}_i$, and a rotating reference frame $K'$, with basis vectors $\mathbf{e}'_i(t)$. Any position vector $\mathbf{x}(t)$ can be represented in either $K$ or $K'$. The relation between velocity $\mathbf{v}$ and acceleration $\mathbf{a}$ in $K$ and the corresponding quantities $\mathbf{v}'$ and $\mathbf{a}'$ in $K'$ is given by

$$\mathbf{a} = \mathbf{a}' + 2\boldsymbol{\omega} \times \mathbf{v}' + \dot{\boldsymbol{\omega}} \times \mathbf{x} + \boldsymbol{\omega} \times (\boldsymbol{\omega} \times \mathbf{x}) \tag{12}$$

$$\mathbf{v} = \mathbf{v}' + \boldsymbol{\omega} \times \mathbf{x}. \tag{13}$$

**Fig. 2** This is a special case of a trajectory for initial values $\mathbf{v}_0$ parallel to $\mathbf{n}$

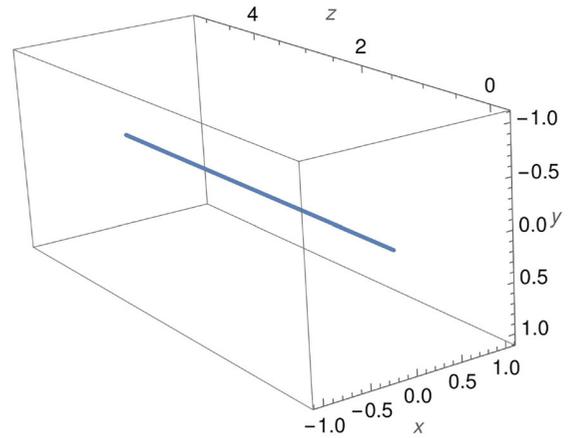

If we set here $\mathbf{v}' = \mathbf{a}' = 0$ (and $\dot{\boldsymbol{\omega}} = 0$), we obtain Eq. (11). The equation of motion (11) for Galilean particles thus simply states that for a point held fixed in $K'$, the sum of the centripetal force and the centrifugal force vanishes.

The general solution of (11) depends on the 9 integration constants $\boldsymbol{\omega}$, $\mathbf{v}_0$, $\mathbf{x}_0$. If the unit vector $\mathbf{n} = \boldsymbol{\omega}/\omega$ parallel to $\boldsymbol{\omega}$ and its magnitude $\omega = |\boldsymbol{\omega}|$ are introduced, the solution takes the form

$$\mathbf{x}(t) = (\mathbf{n} \times \mathbf{v}_0) \frac{1 - \cos \omega t}{\omega} + (\mathbf{v}_0 - \mathbf{v}_0 \cdot \mathbf{n} \, \mathbf{n}) \frac{\sin \omega t}{\omega} + \mathbf{v}_0 \cdot \mathbf{n} \, \mathbf{n} \, t + \mathbf{x}_0. \tag{14}$$

The generic solution, with $\mathbf{v}_0$ neither parallel nor perpendicular to $\mathbf{n}$, is shown in Fig. 1. It has the form of a spiral, i.e., a superposition of a circular motion with a Galilean boost. When $\mathbf{v}_0$ is parallel to $\mathbf{n}$, the circular motion disappears and only the Galilean boost survives, as shown in Fig. 2. We shall come back to this degenerate case in Sect. 5

Finally, to investigate the similarity of the solutions (14) with the Galilean transformations (10), we use the relation

$$\mathbf{r}' := R(\mathbf{n}, \phi)\mathbf{r} = \mathbf{r} - (\mathbf{r} - \mathbf{r} \cdot \mathbf{n} \, \mathbf{n}) \, (1 - \cos \phi) + (\mathbf{n} \times \mathbf{r}) \sin \phi, \tag{15}$$

which describes the effect of a rotation R on a vector $\mathbf{r}$ when the angle of rotation $\phi$ and the axis $\mathbf{n}$ are used as parameters [17]. With the help of (15) and the definitions $\mathbf{r}_0 = \mathbf{v}_0/\omega$, $\mathbf{v}_{\shortparallel} = \mathbf{v}_0 \cdot \mathbf{n} \, \mathbf{n}$, the solution (14) may be written in the form

$$\mathbf{x}(t) = R(\mathbf{n}, \omega t - \pi/2)\mathbf{r}_0 + \mathbf{v}_{\shortparallel} t + \mathbf{n} \times \mathbf{r}_0 - \mathbf{n} \cdot \mathbf{r}_0 \, \mathbf{n} + \mathbf{x}_0. \tag{16}$$

In this formula, all types of transformations belonging to the Galilei group, with the exception of the time translations, occur. We have a rotation of the vector $\mathbf{r}_0$, with an axis $\mathbf{n}$ and an angle $\omega t - \pi/2$, a Galilei boost, with the velocity $\mathbf{v}_{\shortparallel}$, and a spatial translation by $\mathbf{n} \times \mathbf{r}_0 - \mathbf{n} \cdot \mathbf{r}_0 \, \mathbf{n} + \mathbf{x}_0$.

However, the individual group parameters in (16) are complicated functions of the initial conditions $\mathbf{x}_0$, $\mathbf{v}_0$, $\boldsymbol{\omega}$; in particular, the angle of rotation depends on time. Therefore, (16) is still very different from a real Galilei transformation, where all group parameters can be varied independently. As the solutions of (11) do not provide a complete description of of Galilean space we expect that the HLLK will only lead to physically meaningful results if additional restrictions are introduced.

## 4 Transition to Canonical Equations

In order to start the program of the HLLK, we must bring, as a first step, the equations of motion (11) of the Galilean particles into canonical form. For this purpose, the components of the angular velocity may be expressed by Euler angles, which allow to describe the instantaneous state of rotation of the particles. We follow here closely the procedure of Holland [9,18] using a standard notation for the Euler angles $\theta$, $\phi$, $\psi$ [17], which will alternatively be denoted by $\alpha_k$ where $\alpha_1 = \theta$, $\alpha_2 = \phi$, $\alpha_3 = \psi$.

The relation between the time derivatives $\dot{\alpha}_i$ of the Euler angles and the components of the angular velocity is given by

$$\dot{\alpha}_i = A_{ik}\omega_k, \tag{17}$$

where the elements of the matrix $A$ are periodic function of $\theta$ and $\phi$,

$$A = \begin{bmatrix} \cos\phi & \sin\phi & 0 \\ -\cot\theta\sin\phi & \cot\theta\cos\phi & 1 \\ \csc\theta\sin\phi & -\csc\theta\cos\phi & 0 \end{bmatrix}. \tag{18}$$

Using the Euler angles, the original equations of motion (11) now take the form of two second order differential equations,

$$\ddot{\alpha}_l + A_{li}\frac{\partial A_{ik}^{-1}}{\partial\alpha_r}\dot{\alpha}_r\dot{\alpha}_k = 0, \tag{19}$$

$$\ddot{x}_i = \epsilon_{ijk}A_{js}^{-1}\dot{\alpha}_s\dot{x}_k. \tag{20}$$

Note that Eq. (19) only describes the constancy of $\boldsymbol{\omega}$ over time and does not depend on $x$. The equations of motion (19), (20) of the Galilean particles have a somewhat unusual form which stems from the fact that each point of time $t$ is assigned not only a position in $\mathbb{R}^3$ but also a state of rotation, which is described by a point in the group space $SO(3)$ of the rotation group. The 6-dimensional configuration space $\mathbb{R}^3 \times SO(3)$ agrees with that of the rigid body, as already mentioned by Holland [9,18].

The Eqs. (19), (20) can easily be derived, by means of a variation with respect to $x_k$ and $\alpha_k$, from a Lagrangian $L$ of the form

$$L(x, \alpha, \dot{x}, \dot{\alpha}) = d_0 A_{ik}^{-1}(\alpha)\dot{x}_i\dot{\alpha}_k, \tag{21}$$

if one observes that the matrix elements of $A$ fulfill the relations

$$\left[\frac{\partial A_{kr}^{-1}}{\partial\alpha_s} - \frac{\partial A_{ks}^{-1}}{\partial\alpha_r}\right]A_{si} = \epsilon_{ijk}A_{jr}^{-1}. \tag{22}$$

To prove this relation, we use the definition (18) and verify the equality of the two sides of (22) with the help of the computer algebra program Mathematica.

Before we construct the canonical equations, we note the following inconspicuous but important point. We introduced in Eq. (21) a constant $d_0$ as a pre-factor. Neither its numerical value nor its dimension is determined by the form of the equations of motion. We can choose both freely, at least at this point. However, due to the projection onto the configuration space, to be performed later in Sect. 6, a new fundamental constant, say $c$, will be generated in our theory. This seems to be a generic property of the HLLK (In the earlier case of massive particles studied in I–IV, Planck's quantum of action was generated in this way). Clearly, the choice of the constant $d_0$ is important because it determines this new constant $c$. In a sense, we are even able to *derive* the constant $c$, to the extent that we are able to justify our choice of $d_0$ sufficiently well. Of course, this derivation can not concern the numerical value of $c$, this value will be determined by nature alone. However, by means of a physically justified determination of the *dimension* of $d_0$ we can at least derive the dimension of $c$, which already gives us a crucial information about its physical meaning. A reasonable choice for the dimension of the Lagrangian $L$ seems to be *energy*, the same dimension as in the theory of particles with non-zero mass. We have already chosen the designation (Galilean) "particles" for the objects, whose trajectories are determined by the solutions of Eq. (11). However, this was just a semantic assignment. If we now determine that $d_0$ has the dimension $g\,cm$ (which gives $L$ the dimension of an energy) then we make a formal determination of these objects as particles—which goes beyond the previous semantic assignment. In Sect. 6.2, we will see that this choice of $d_0$ implies the dimension of a velocity for $c$.

The canonical equations may be constructed using the standard method. The momentum $p_k$, canonically conjugate to $x_k$, is given by

$$p_i := \frac{\partial L}{\partial \dot{x}_i} = d_0 A_{ik}^{-1}(\alpha)\dot{\alpha}_k. \tag{23}$$

Equation (17) shows that $p_i = d_0\omega_i$. Since $\dot{\omega}_i = 0$, the momentum $p_i$ is conserved. Thus, Galilean particles are subject to an unusual, strongly degenerate dynamics. The momentum $\pi_i$ canonically conjugate to $\alpha_i$ is given by

$$\pi_i := \frac{\partial L}{\partial \dot{\alpha}_i} = d_0 A_{ki}^{-1}(\alpha)\dot{x}_k. \tag{24}$$

The momentum $p_k$ does not depend on $\dot{x}_k$ but only on $\dot{\alpha}_k$; an analogous behavior is shown by $\pi_k$. Hamilton's function,

$$H(x, \alpha, p, \pi) = d_0^{-1} A_{ki}(\alpha) p_i \pi_k \tag{25}$$

does not depend on $x$, in agreement with the fact that $p$ does not depend on time. The deeper reason for the independence of $L(x, \alpha, \dot{x}, \dot{\alpha})$ and $H(x, \alpha, p, \pi)$ from $x$ is the translation-invariance of the basic equation of motion (11). The canonical equations are given by

$$\dot{x}_k = d_0^{-1} A_{ik}(\alpha)\pi_i, \qquad \dot{\alpha}_k = d_0^{-1} A_{ki}(\alpha)p_i, \tag{26}$$

$$\dot{p}_k = 0, \qquad\qquad \dot{\pi}_k = -d_0^{-1} \frac{\partial A_{ji}(\alpha)}{\partial \alpha_k} p_i \pi_j. \tag{27}$$

The solutions of the ordinary differential equations (26),(27) are trajectories in a 12-dimensional phase space $\boldsymbol{\Omega}$ [cotangent bundle of the rigid body configuration space $\mathbb{R}^3 \times SO(3)$], which is spanned by the coordinates $x, \alpha, p, \pi$. Remarkably, the velocity $\dot{x}$ is not proportional to the momentum $p$. Another interesting variable, defined by its components

$$m_k = A_{ik}(\alpha)\pi_i, \tag{28}$$

takes the dimension of an angular momentum (if the parameter $d_0$ is defined as in Sect. 4). The elements of the matrix $A$ obey the relations $A_{rk}\partial_r A_{ji} - A_{ri}\partial_r A_{jk} = \epsilon_{ikr} A_{jr}$. Using these identities, it is easy to see that the $m_k$ fulfill also the Poisson bracket relations

$$\{m_i, m_k\} = \epsilon_{ikl} m_l \tag{29}$$

characterizing angular momentum components, as noted already by Holland [9,18]. Remarkably, according to the equations of motion [see the left member of (26)], the velocity $\vec{v} = \dot{\vec{x}}$ agrees for arbitrary times with the angular momentum $\mathbf{m}$, apart from a proportionality constant. The Hamiltonian function may consequently be written in the form $H = v_k p_k$.

In order to get rid of the ugly prefactors $d_0^{-1}$, the position variables $x_i$ may be replaced by the new variables

$$u_i = d_0 x_i, \tag{30}$$

which have dimension $g\,cm^2$. The new momentum variables, canonically conjugate to $u_i$ are then given by $d_0^{-1} p_i = \omega_i$. Using the new variables $u_i, \omega_i$, the Lagrange function and the Hamilton function take the form

$$L(u, \alpha, \dot{u}, \dot{\alpha}) = A_{ik}^{-1}(\alpha)\dot{u}_i\dot{\alpha}_k, \tag{31}$$

$$H(u, \alpha, \omega, \pi) = A_{ki}(\alpha)\omega_i\pi_k, \tag{32}$$

and the canonical equations are given by

$$\dot{u}_k = A_{ik}(\alpha)\pi_i, \qquad \dot{\alpha}_k = A_{ki}(\alpha)\omega_i, \tag{33}$$

$$\dot{\omega}_k = 0, \qquad\qquad \dot{\pi}_k = -\frac{\partial A_{ji}(\alpha)}{\partial \alpha_k} \omega_i \pi_j. \tag{34}$$

We will change freely between Eqs. (32)–(34) and the original equations (25)–(27) where the important parameter $d_0$ appears explicitly.

The set of all solutions, together with a probability (or density) distribution at $t = 0$, defines a statistical (or particle) ensemble formulated by means of Lagrangian coordinates. The next step in the formulation of the HLLK is the transition to the usual (Eulerian) formulation, which describes physical properties by means of *fields* depending on the coordinates of a suitable space. In our case, this space is $\mathbf{\Omega}$.

## 5 Lagrangian to Eulerian transition

For the general considerations of this section, it is convenient to use the original equations of motion (26),(27). We write the solutions in the form

$$x_k(t) = x_k(t, x^0, p^0, \alpha^0, \pi^0), \tag{35}$$
$$p_k(t) = p_k(t, x^0, p^0, \alpha^0, \pi^0), \tag{36}$$
$$\alpha_k(t) = \alpha_k(t, x^0, p^0, \alpha^0, \pi^0), \tag{37}$$
$$\pi_k(t) = \pi_k(t, x^0, p^0, \alpha^0, \pi^0), \tag{38}$$

where $x_k^0, \alpha_k^0, p_k^0, \pi_k^0$ are the initial values of the dynamic variables at $t = 0$. If these equations can be solved for $x_k^0, \alpha_k^0, p_k^0, \pi_k^0$, then they define a *flow* in $\mathbf{\Omega}$, that is, there is a mapping from $\mathbf{\Omega}$ onto itself at each instant of time $t$.

In order to discuss the physical meaning of this mapping, it is convenient to abandon the special meaning of $H$ and to substitute an arbitrary observable $A(Q, P)$ in place of $H$. We think of these observables, which depend on $2n$ phase space variables $Q, P$, as important physical quantities like energy, momentum or angular momentum (all defined by a fundamental symmetry of space-time). Each $A(Q, P)$ defines canonical equations, whose solutions (if invertibly) represent a group of transformations, each element of which maps the phase space onto itself. The observables $A$ represent the classical equivalent of the associated Hermitian operators $\hat{A}$ in QT. The one-parameter group of transformations generated by $A$ corresponds in QT to the one-parameter group of unitary operators generated by $\hat{A}$. These relations, basically known for a long time [19], have recently been used by the author as an essential component in a reconstruction of QT, see II.

In the present work, we assume that analogous relations hold also for massless particles. Accordingly, $A$ is identified with the Hamiltonian function (25) of Galilean space-time. The variables $Q$ and $P$ are identified with the configuration space variables $x, \alpha$ and the corresponding canonically conjugate momenta $p, \pi$, respectively. We first assume that the mapping defined by (35)-(38), from $\mathbf{\Omega}$ onto itself, is $1 : 1$, but analyze this question more carefully at the end of this section.

The Lagrangian to Eulerian transition, which we perform now, is a straightforward generalization of the well-known method in fluid mechanics (where $n = 3$) to arbitrary $n$. The details were reported in II. From the canonical equations, it follows that an arbitrary density $\rho(x, \alpha, p, \pi, t)$ obeys the Liouville equation

$$\frac{\partial \rho}{\partial t} + \frac{\partial \rho}{\partial x_k} \frac{\partial H}{\partial p_k} + \frac{\partial \rho}{\partial \alpha_k} \frac{\partial H}{\partial \pi_k} - \frac{\partial \rho}{\partial p_k} \frac{\partial H}{\partial x_k} - \frac{\partial \rho}{\partial \pi_k} \frac{\partial H}{\partial \alpha_k} = 0, \tag{39}$$

During the reconstruction of QT in I–IV, an analogous quantity $\rho(x, p, t)$ was assigned the physical meaning of a probability density. In contrast to this, we do not specify the physical meaning of the density $\rho(x, \alpha, p, \pi, t)$, at least at the moment. We assume further, as in I–IV, that the dynamical variable $\rho(x, \alpha, p, \pi, t)$ is not sufficient to perform the transition to a *quantum theory*. As a second variable we choose, again in analogy to I–IV, the variable $S(x, \alpha, p, \pi, t)$, defined by

$$S(x^0, \alpha^0, p^0, \pi^0, t) = \int^t du \left[ p_k \frac{\partial H}{\partial p_k} + \pi_k \frac{\partial H}{\partial \pi_k} - H \right]. \tag{40}$$

The action variable $S(x, \alpha, p, \pi, t)$ obeys the following differential equation, which is referred to as action equation,

$$\frac{\partial S}{\partial t} + \frac{\partial S}{\partial x_k}\frac{\partial H}{\partial p_k} + \frac{\partial S}{\partial \alpha_k}\frac{\partial H}{\partial \pi_k} - \frac{\partial S}{\partial p_k}\frac{\partial H}{\partial x_k} - \frac{\partial S}{\partial \pi_k}\frac{\partial H}{\partial \alpha_k} =$$
$$p_k\frac{\partial H}{\partial p_k} + \pi_k\frac{\partial H}{\partial \pi_k} - H. \qquad (41)$$

We can combine Eqs. (39), (41) into a single differential equation using the complex-valued field variable $\psi(x, \alpha, p, \pi, t)$, defined by

$$\psi = \sqrt{\rho}\exp\left(\frac{\iota}{s_0}S\right). \qquad (42)$$

Relation (42) is sometimes referred to as a "linearizing transformation". It was originally introduced by Madelung in a three-dimensional context in order to bring the Schrödinger equation into a "hydrodynamic form" [20]. In this formula, we have introduced a parameter $s_0$ which has the dimension of an action and has no physical meaning at all (it is part of the definition of the quantity $\psi$ and may take any numerical value).

Using the relations (39), (41) and (42), we see that all derivatives of $\rho$ and $S$ can be expressed by derivatives of $\psi$, so that (39), (41) can be replaced by the single differential equation

$$\frac{s_0}{\iota}\frac{\partial \psi}{\partial t} - \frac{\partial H}{\partial p_k}\left(p_k - \frac{s_0}{\iota}\frac{\partial}{\partial x_k}\right)\psi - \frac{\partial H}{\partial \pi_k}\left(\pi_k - \frac{s_0}{\iota}\frac{\partial}{\partial \alpha_k}\right)\psi$$
$$-\frac{s_0}{\iota}\left(\frac{\partial H}{\partial x_k}\frac{\partial}{\partial p_k} + \frac{\partial H}{\partial \alpha_k}\frac{\partial}{\partial \pi_k}\right)\psi + H\psi = 0, \qquad (43)$$

for the complex variable $\psi$. In this field-theoretic form of the equations of motion (26), (27), the projection onto the configuration space $\mathbb{R}^3 \times SO(3)$ becomes particularly simple, as will be shown in the next section.

We now return to the question whether the mapping defined by the solutions of the equations of motion is $1:1$. We use the equations of motion in the form (33),(34); the mapping has the form of Eqs. (35)-(38), with $x$ and $p$ replaced by $u$ and $\omega$. A point in $\mathbf{\Omega}$ is denoted by $x$, its components are denoted by $x_\mu$, where $(x_\mu) = (u, \omega, \alpha, \pi)$, $\mu = 1, .., 12$. The initial values at time $t = 0$ are denoted by $y$, its components $y_\mu$ are given by $(y_\mu) = (u^0, \omega^0, \alpha^0, \pi^0)$. The mapping can then be written in the compact form $x = f(t, y)$, and this mapping will be $1:1$ if the determinant of the functional matrix

$$f' = \begin{bmatrix} \partial_1 f_1 & \partial_2 f_1 & \cdots & \partial_{12} f_1 \\ \vdots & \vdots & \ddots & \vdots \\ \partial_1 f_{12} & \partial_2 f_{12} & \cdots & \partial_{12} f_{12} \end{bmatrix} \qquad (44)$$

(where $\partial_\mu$ denotes the derivative with respect to $y_\mu$) is different from 0 everywhere in $\mathbf{\Omega}$. We are unable to perform a complete analysis of this determinant as the time-dependence of the Euler angles is very complicated [21], even for the simple case of constant angular velocity considered here. We will examine only the case where the velocity $\mathbf{v}$ is parallel to $\boldsymbol{\omega}$ (this is the special case shown in Fig. 2). This means $v_k := \dot{u}_k = r\omega_k$ where $r$ is a real number. It is easy to see that in this special case the determinant of the functional matrix (44) vanishes.

To show this, we first assume that the parallelism between $v_k$ and $\omega_k$ holds true only at $t = 0$. Thus, the initial values obey the relations $v_k^0 = r\omega_k^0$. We know that $\boldsymbol{\omega}$ is constant, thus $\omega_k = \omega_k^0$ in agreement with the left member of Eq. (34). Equation (11) shows that all derivatives of $\mathbf{v}$ with respect to $t$ vanish at $t = 0$. It follows that the parallelism $v_k^0 = r\omega_k^0$ at $t = 0$ implies the parallelism $v_k = r\omega_k$ for arbitrary $t$. The left member of Eq. (33) implies $A_{ik}(\alpha^0)\pi_i^0 = r\omega_k^0$. Thus, the initial values of $\pi_i$ can be expressed by the initial values of $\alpha$ and $\omega$ using the relation $\pi_l^0 = r A_{kl}^{-1}(\alpha^0)\omega_k^0$. As a consequence, the last three columns of the functional matrix (44) vanish, since nowhere in the solutions do the variables $\pi_l^0$ appear any more. For the solutions obeying

$$\mathbf{v} = r\boldsymbol{\omega}, \qquad (45)$$

the Lagrangian to Eulerian transition cannot be carried out. These solutions must be excluded by an appropriate additional constraint. This will be done in Sect. 7.3.

An important part of HLLK is the projection from phase space $\boldsymbol{\Omega}$ onto $\mathbb{R}^3 \times SO(3)$. This step is similar to the usual quantization rule, in that the canonical momenta are replaced by operators acting on configuration space. This projection, which will be carried out in Sect. 6, is, in the case of massless particles treated here, not sufficient to generate a physically meaningful field theory. In order to arrive at a field theory with independent variables $t$, $x_k$, a further reduction from $\mathbb{R}^3 \times SO(3)$ to Euclidean configuration space $\mathbb{R}^3$ must be performed. This step will be reported in Sect. 7.

## 6 Projection to rigid body configuration space

From now on, we use the original system of Eqs. (26),(27) for the variables $x_k$, $\alpha_k$ and the canonically conjugate momenta $p_k$, $\pi_k$. Our starting point is the linearized equation of motion (43) for $\psi$, with the Hamiltonian function (25). In Eq. (43), two constant parameters $d_0$ and $s_0$ occur, which have—as part of the *definition* of the Lagrangian function and the dynamical variable $\psi$—no physical meaning at this time.

### 6.1 The projection

The projection onto $\mathbb{R}^3 \times SO(3)$ is done exactly as in I, II. One replaces $\psi(x, \alpha, p, \pi, t)$ by an unknown function $\psi(x, \alpha, t)$ denoted for simplicity by the same symbol. Thus, the third bracket in (43) disappears. To remove $p_k$ and $\pi_k$ from the operator expression, the only remaining option is to perform the following substitution by operators:

$$p_k \Rightarrow \hat{p}_k := \frac{s_0}{\iota} \frac{\partial}{\partial x_k}, \qquad \pi_k \Rightarrow \hat{\pi}_k := \frac{s_0}{\iota} \frac{\partial}{\partial \alpha_k}. \tag{46}$$

This makes the first two brackets disappear and the Hamiltonian function $H(x, \alpha, p, \pi)$ becomes an operator $\hat{H} = H(x, \alpha, \hat{p}, \hat{\pi})$. So the well-known quantization recipe finds a natural explanation as a projection rule. We mention that the standard quantization recipe (without the $\alpha$-degree of freedom) has also been used to derive Maxwell's equations from the relativistic particle equations of motion [22]. The projection transforms Eq. (43) into the 'Schrödinger-like' differential equation

$$\left( \frac{s_0}{\iota} \frac{\partial}{\partial t} + \hat{H} \right) \psi = 0. \tag{47}$$

The Hamilton operator now has the form $\hat{H} = d_0^{-1} A_{ki}(\alpha) \hat{p}_i \hat{\pi}_k$ if we choose the order of the operators according to the non-quantized Hamilton function (25). However, the exact form of $\hat{H}$ is not yet known, because the order of the non-commuting quantities $A_{ki}(\alpha)$ and $\hat{\pi}$ has still to be fixed. (There is no ordering problem between $x$ and $\hat{p}$ because $\hat{H}$ does not depend on $x$). The standard method of symmetrization cannot be applied here because of the non-Euclidean configuration space $\mathbb{R}^3 \times SO(3)$.

To clarify the ordering problem, we use the inner product

$$(\psi, \phi) = \int_\Sigma \mathrm{d}\sigma \, \psi^*(x, \alpha, t) \phi(x, \alpha, t), \tag{48}$$

where the abbreviations $\Sigma = \mathbb{R}^3 \times SO(3)$, $\mathrm{d}\sigma = \mathrm{d}^3 x \, \mathrm{d}^3 \alpha \, \sin \alpha_1$ were used and the Euler angles vary according to $\alpha_1 \in [0, \pi]$, $\alpha_2 \in [0, 2\pi]$ $\alpha_3 \in [0, 2\pi]$. One can now show that the Hamilton operator

$$\hat{H} = d_0^{-1} A_{ki}(\alpha) \hat{p}_i \hat{\pi}_k \tag{49}$$

is Hermitian with respect to the scalar product (48) as long as one restricts oneself to states in the space of $2\pi$-periodic functions. The proof is given in "Appendix A". The reverse order of $A_{ki}(\alpha)$ and $\hat{\pi}_k$ can be excluded.

As a consequence of the quantization, the angular momentum components $m_k$, defined by (28), become operators

$$\hat{m}_k = A_{ik}(\alpha) \hat{\pi}_i = \frac{s_0}{\iota} A_{ik}(\alpha) \frac{\partial}{\partial \alpha_i}, \tag{50}$$

The $m_k$ satisfy the commutation relations ($[\hat{a}, \hat{o}] = \hat{a}\hat{o} - \hat{o}\hat{a}$)

$$\left[\hat{m}_i, \hat{m}_k\right] = -\frac{s_0}{\iota}\epsilon_{ikl}\hat{m}_l. \tag{51}$$

In the proof of (51), the same identity is used as in the proof of the corresponding Poisson bracket relation (29). Using the operators $\hat{m}$, the Hamilton operator takes the form $\hat{H} = d_0^{-1}\hat{m}_k\hat{p}_k$. It is then sometimes interpreted as helicity, a projection of angular momentum (or spin) onto momentum.

The velocity components $v_k$ become operators $\hat{v}_k = d_0^{-1}\hat{m}_k$ due to quantization. When $\hat{v}$ is used, $\hat{H}$ takes the form $\hat{H} = \hat{v}_k\hat{p}_k$, and thus becomes formally similar to the Hamilton operator of Dirac's theory in the limiting case of vanishing mass. Of course, in Dirac's theory, the velocity operator is given by $c$ times $\alpha$, where $c$ is the speed of light, and $\alpha$ is a three-vector whose components are $4 \times 4$ matrices. In the present case, the velocity is still a differential operator, but we will perform, in Sect. 7, a second projection which maps the velocity components to matrices.

## 6.2 A new fundamental constant

If we insert the definitions of $\hat{p}_k$ and $\hat{\pi}_k$, and multiply with a factor $s_0^{-1}$, the Schrödinger-like equation (47) takes the form

$$\left(\frac{1}{\iota}\frac{\partial}{\partial t} - d_0^{-1}s_0\,A_{ki}(\alpha)\frac{\partial}{\partial\alpha_k}\frac{\partial}{\partial x_i}\right)\psi = 0. \tag{52}$$

Thus, the constants $d_0$ and $s_0$ do no longer occur individually, but only in the combination $d_0^{-1}s_0$, which has the dimension of a velocity. While the quantities $d_0$ and $s_0$ were originally introduced as purely mathematical definitions, now, after the projection, they can no longer be changed at will. Thus, a new natural constant $d_0^{-1}s_0$ with the dimension of a velocity appears, which we denote by $c$. Experiments show that the constant $c$ must be assigned the numerical value of the speed of light. As is well-known, the appearance $c$ of has dramatic consequences for the structure of spacetime.

During the quantization of massive particles, reported in I–IV, the parameter $s_0$ was identified with Planck's quantum of action $\hbar$. This constant has the physical meaning of an accuracy limit for measurements, and is in this sense characteristic for QT. The absence of $\hbar$ in (52) indicates that this equation should not be interpreted in the same probabilistic sense as Schrödinger's equation for massive particles. We discuss this point in more detail in Sect. 8.

Although the constants $d_0$ and $s_0$ only occur in the combination $d_0^{-1}s_0$, fundamental relations between $d_0$, $s_0$ on the one hand and $d_0^{-1}s_0$ on the other can be determined. This is possible because $s_0$ and $d_0$ were both introduced in a well-founded physical context. The quantity $s_0$ was introduced as a unit of measurement of the classical action, and it seems natural to assume a connection with QT here and thus to identify the numerical value of $s_0$ with the value of Planck's quantum of action $\hbar$. The quantity $d_0$ was introduced as a pre-factor of a Lagrange function that describes "Galilean equations of motion". The dimension of $d_0$ was determined under the assumption that the objects obeying these Galilean equations of motion have the character of particles (i.e., the Lagrange function was assigned the dimension of an energy). This means that the two concepts that "in the background" establish the dimensions of the quantities $s_0$ and $d_0$ are those of QT and the point particle.

The dimensions $s_0$ and $d_0$ chosen in this way are compatible with the fact that the new natural constant $d_0^{-1}s_0$ occurring in Eq. (52) has the dimension of a velocity. Obviously, only two of the three constants $s_0$, $d_0$ and $d_0^{-1}s_0$ are independent of each other, and for the assignment $s_0 = \hbar$, $d_0^{-1}s_0 = c$, $d_0$ is given by

$$d_0 = \frac{\hbar}{c}, \tag{53}$$

as the ratio of two fundamental constants of nature which characterize the transition from classical physics to QT and to special relativity, respectively. We may also say that only two of the following three concepts or ideas are independent of each other:

1. The idea of a fundamental limitation of measurement accuracy, as expressed by the occurrence of the constant $\hbar$ in QT.
2. The idea of a fundamental limitation of the speed of propagation as expressed by the occurrence of the constant $c$ in the special theory of relativity.
3. The idea of massless particles (photons) as a means of describing empty space.

This relationship is also responsible for the famous linear relationship between momentum and angular frequency, which we derive next.

The constant $d_0$ itself does not occur in the field equation (52), but it occurs in the definition of the canonical momentum $p_k = d_0 \omega_k$ [see text following Eq. (23)]. Accordingly, the linear relationship $cp_k = \hbar \omega_k$ exists between the components $p_k$ of the canonical momentum of the Galilean particle and the angular velocity, or angular frequency, $\omega_k$. The corresponding relation between the absolute values $p$ and $\omega$ is given by

$$cp = \hbar \omega. \tag{54}$$

Equation (54) establishes a relation between particle properties and wave properties, as does the related equation $p = h/\lambda$ between momentum $p$ and wavelength $\lambda$. These famous relations, associated with the names of Einstein and De Broglie, were, remarkably, discovered very early in the development of QT [23]. We have found here, using the framework of HLLK, a new derivation of these relations.

## 7 Projection to Euclidean configuration space

We perform, as a first step of the projection to $\mathbb{R}^3$, a general expansion in the parameter space of the rotation group SO(3). A suitable set of basis functions is provided by the so-called Wigner D-matrices [24]

$$D^l_{m,n}(\alpha) = e^{\imath m \alpha_2} d^l_{m,n}(\cos \alpha_1) e^{\imath n \alpha_3}, \tag{55}$$

where the $d^l_{m,n}$ are generalized associated Legendre functions. The $D^l_{m,n}$ are the matrix elements of the rotation operator in a $l, m$ (angular momentum) basis ($l = 0, 1, 2, .., m = -l, .., +l$). For each $l$, the $(2l + 1) \times (2l + 1)$ matrices $D^l_{m,n}$ represent an irreducible representation of $SO(3)$. The homomorphism property may be expressed in the form

$$D^l_{m,n}(g_2 g_1) = \sum_{m'=-l}^{l} D^l_{m,m'}(g_2) D^l_{m',n}(g_1), \tag{56}$$

where $g_2 g_1$ is the product of the rotations $g_1, g_2 \in SO(3)$. The $D^l_{m,n}$ form an orthonormalized set of basis functions on $SO(3)$,

$$\left( D^l_{m,n}, D^r_{p,q} \right)_R = \delta_{lr} \delta_{mp} \delta_{nq}, \tag{57}$$

with the inner product being defined by

$$(\varphi, \chi)_R = \int_{SO(3)} d^3 \alpha \sin \alpha_1 \varphi^*(\alpha) \chi(\alpha), \tag{58}$$

Furthermore, one can show that this set is complete. Our wave function may consequently be written in the form

$$\psi(x, \alpha, t) = \sum_{l=0}^{\infty} \sum_{m=-l}^{l} \sum_{n=-l}^{l} F^l_{m,n}(x, t) D^l_{m,n}(\alpha). \tag{59}$$

This expansion shows that $\psi$ transforms under rotations according to a reducible representation, namely the direct sum of all irreducible representations. If we were to project Eq. (52) onto $\mathbb{R}^3$ without any further constraints, we would have to introduce a variable with infinitely many components. This is of course not acceptable. We need a variable with a relatively small number of components which, more importantly, transforms according to an irreducible representation of the rotation group. Thus, the irreducibility postulate used so successfully by Wigner plays a decisive role in the projection from $\mathbb{R}^3 \times SO(3)$ to $\mathbb{R}^3$.

### 7.1 Reduction to $l = 1$

We choose the irreducible representation $l = 1$. As a basis in the corresponding subspace, we use the three functions $D^1_{a,0}(\alpha), a = -1, 0, 1$; these transform according to the $l = 1$ representation, as shown by Eq. (56). In the following, we write $D_a$ instead of $D^1_{a,0}$ omitting the indices $1, 0$. Then the expansion (59) reduces to

$$\psi(x, \alpha, t) = \sum_{a=-1}^{1} F_a(x, t) D_a(\alpha), \tag{60}$$

where the three orthonormalized basis functions are given by

$$D_{-1}(\alpha) = \iota \frac{\sqrt{3}}{4\pi} e^{-\iota \alpha_2} \sin \alpha_1, \tag{61}$$

$$D_0(\alpha) = \frac{\sqrt{3}}{2\pi \sqrt{2}} \cos \alpha_1, \tag{62}$$

$$D_1(\alpha) = \iota \frac{\sqrt{3}}{4\pi} e^{\iota \alpha_2} \sin \alpha_1. \tag{63}$$

The projection to the three-dimensional subspace generates a theory in the Euclidean configuration space $\mathbb{R}^3$ (a conventional field theory) for a wave function $\psi(x, t)$ with the three components $F_{-1}(x, t)$, $F_0(x, t)$, $F_1(x, t)$ (indices $a, b$ running from $-1$ to $1$). The Schrödinger-like differential equation (47) takes the form

$$\frac{\hbar}{\iota} \frac{\partial F_a}{\partial t} + d_0^{-1} \hat{p}_i M^i_{a,b} F_b = 0, \tag{64}$$

where the quantities $M^i_{a,b}$ are the matrix elements of the operators $\hat{m}_i$ in the $l = 1$ basis,

$$M^i_{a,b} = \left( D_a, \hat{m}_i D_b \right)_R, \tag{65}$$

$$M^1 = \frac{\hbar}{\sqrt{2}} \begin{bmatrix} 0 & 1 & 0 \\ 1 & 0 & 1 \\ 0 & 1 & 0 \end{bmatrix}, \quad M^2 = \frac{\hbar}{\sqrt{2}} \begin{bmatrix} 0 & \iota & 0 \\ -\iota & 0 & \iota \\ 0 & -\iota & 0 \end{bmatrix}, \quad M^3 = \hbar \begin{bmatrix} -1 & 0 & 0 \\ 0 & 0 & 0 \\ 0 & 0 & 1 \end{bmatrix}. \tag{66}$$

Performing a necessary rearrangement of indices, the matrices $M^i$ coincide with the matrices $J_i$ derived by Holland from Maxwell's equations [9].

The matrices $M^i$ fulfill the same commutation relations

$$\left[ M^i, M^k \right] = -\frac{\hbar}{\iota} \epsilon_{ikl} M^l, \tag{67}$$

as the differential operators $\hat{m}_k$ from which they were derived. The constant matrices $M^i$ are angular momentum operators (for $l = 1$), but their action consists only in a mixing of the three components of $\psi$. Thus, they do not describe an orbital angular momentum but an *internal* degree of freedom called spin (spin 1). Based on the present derivation, it is clear that the rotational motion associated with this internal degree of freedom must not be interpreted as rotation of individual massless particles (photons), but as collective rotational motion in a (statistical ?) ensemble of massless particles.

The components $\hat{v}_k$ of the velocity operator are transformed by the projection into the $3 \times 3$ matrices $V^k = d_0^{-1} M^k$. We note that both the velocity $V^k$ and the angular momentum $M^k$ describe internal degrees of freedom. Using the three-component wave function $\psi$, Eq. (64) may be written in the more compact form

$$\left( \frac{\hbar}{\iota} \frac{\partial}{\partial t} + V^i \hat{p}_i \right) \psi = 0, \tag{68}$$

which shows the formal similarity with the massless Dirac theory particularly clearly.

## 7.2 Transition to Cartesian basis

If we assume that the theory described by (68) makes physical sense, then it seems useful to replace the spherical coordinates with the more commonly used Cartesian coordinates. In concrete terms, this means that we have to look for a representation of the spin matrices that corresponds to the transformation behavior of real vector fields in $\mathbb{R}^3$. These are given by $S_{jk}^i = \frac{\hbar}{\iota}\epsilon_{ijk}$, or

$$S^1 = \frac{\hbar}{\iota}\begin{bmatrix} 0 & 0 & 0 \\ 0 & 0 & 1 \\ 0 & -1 & 0 \end{bmatrix}, \quad S^2 = \frac{\hbar}{\iota}\begin{bmatrix} 0 & 0 & -1 \\ 0 & 0 & 0 \\ 1 & 0 & 0 \end{bmatrix}, \quad S^3 = \frac{\hbar}{\iota}\begin{bmatrix} 0 & 1 & 0 \\ -1 & 0 & 0 \\ 0 & 0 & 0 \end{bmatrix}. \tag{69}$$

As for the prefactors $\frac{\hbar}{\iota}$, these were added in order to be able to use the classical result as part of QT; of course, the classical rotation operator itself does not depend on $\hbar$.

The transition to the Cartesian basis is performed by means of the transformation $F_a = U_{ai}G_i$, $S_{nm}^i = U_{na}^{-1}M_{ab}^i U_{bm}$, where indices $a, b$ take values from $-1, 0, 1$, all other indices running from 1 to 3, and the unitary matrix $U$ is given by

$$U = \frac{1}{\sqrt{2}}\begin{bmatrix} 1 & \iota & 0 \\ 0 & 0 & \sqrt{2} \\ -1 & \iota & 0 \end{bmatrix}. \tag{70}$$

In terms of the new Cartesian field components $G_i$ and spin matrices $S^i$ the Schrödinger-like differential equation (64) takes the form

$$\frac{\hbar}{\iota}\frac{\partial G_n}{\partial t} + d_0^{-1}\hat{p}_i S_{n,l}^i G_l = 0. \tag{71}$$

Denoting the components $d_0^{-1}S^i$ of the velocity operator with the same symbol $V^i$ as before, and using also the same symbol $\psi$ for the complex quantity with components $G_1, G_2, G_3$, Eq. (71) may also be written in the more compact form (68).

If we now introduce real vector fields **E** and **B** with components defined by $G_k = E_k + \iota B_k$, it follows immediately from (71) that these fields satisfy the two time-dependent electromagnetic field equations

$$\frac{\partial}{\partial \mathbf{r}} \times \mathbf{B} - \frac{1}{c}\frac{\partial \mathbf{E}}{\partial t} = 0, \tag{72}$$

$$\frac{\partial}{\partial \mathbf{r}} \times \mathbf{E} + \frac{1}{c}\frac{\partial \mathbf{B}}{\partial t} = 0. \tag{73}$$

We have not yet used the constraint that states corresponding to Eq. (45) are not allowed. Next, we try to derive the two remaining Maxwell equations from this constraint.

## 7.3 Derivation of transversality conditions

The additional condition we have to take into account was formulated in Sect. 5, in the context of the description of particle orbits in phase space. It stated that those orbits which fulfil the condition $\mathbf{v} = r\boldsymbol{\omega}$ [see Eq. (45)] are to be excluded from consideration. If the relation $\mathbf{p} = d_0\boldsymbol{\omega}$ [see text following Eq. (23)] is used, then the additional condition means that the velocity $\mathbf{v}$ must not be parallel to the direction of propagation $\mathbf{n}(p) = \mathbf{p}/p$, where $p = \sqrt{p_1^2 + p_2^2 + p_3^2}$.

We have made three drastic changes to the original picture of the motion of (massless) particles in the course of the present development. In the first step, we moved from particle coordinates to space coordinates, replacing ordinary with partial differential equations. As a consequence, the concept of trajectories lost its meaning; it was replaced by the concept of continuous distributions in phase space. In the second step, we moved from phase space to rigid body configuration space. In this step the transition to QT was carried out—in the sense that observables in phase space were replaced by operators in configuration space. Remarkably, with this single step, we left not only the realm of classical physics but also the realm of non-relativistic physics. In the third step, the projection onto the Euclidean configuration space was performed. The number of degrees of freedom was further drastically reduced by replacing operators on $SO(3)$ by $3 \times 3$ matrices.

Despite these major changes, the original condition (45) can easily be "translated" into the present version of the theory. Let us consider the operators of velocity $V^i$ and momentum $\hat{p}_i$, replacing the original quantities $v_i$ and $p_i$. The eigenvalues of the components

$$V^1 = \frac{c}{\iota} \begin{bmatrix} 0 & 0 & 0 \\ 0 & 0 & 1 \\ 0 & -1 & 0 \end{bmatrix}, \quad V^2 = \frac{c}{\iota} \begin{bmatrix} 0 & 0 & -1 \\ 0 & 0 & 0 \\ 1 & 0 & 0 \end{bmatrix}, \quad V^3 = \frac{c}{\iota} \begin{bmatrix} 0 & 1 & 0 \\ -1 & 0 & 0 \\ 0 & 0 & 0 \end{bmatrix}. \tag{74}$$

of the velocity operator are $0, \pm c$. The particles that form the quantum mechanical counterpart to the original Galilean particles—we can call them photons—are thus either at rest or moving at the speed of light. This means that the photons that make up our ensemble may have unusual kinematics; for example, the velocity is always parallel to the spin as shown by the relation $V^i = d_0^{-1} S^i$.

In order to find the possible values of $V$ relative to the direction of propagation $\mathbf{n}(p)$, we solve the eigenvalue problem of the operator $V^i \hat{p}_i$ [25]:

$$\begin{bmatrix} E & \iota c p_3 & -\iota c p_2 \\ -\iota c p_3 & E & \iota c p_1 \\ \iota c p_2 & -\iota c p_1 & E \end{bmatrix} \begin{pmatrix} u_1 \\ u_2 \\ u_3 \end{pmatrix} = 0. \tag{75}$$

We obtain three eigenvalues $E^a$ ($a = 0, +, -$) given by

$$E^0 = 0, \quad E^\pm = \pm c p, \tag{76}$$

with the corresponding orthonormalized eigenvectors

$$u^0(p) = \frac{1}{p} \begin{pmatrix} p_1 \\ p_2 \\ p_3 \end{pmatrix}, \quad u^\pm(p) = \frac{1}{\sqrt{2 p^2 \left( p_1^2 + p_2^2 \right)}} \begin{pmatrix} \pm \iota p_2 p - p_1 p_3 \\ \mp \iota p_1 p - p_2 p_3 \\ p_1^2 + p_2^2 \end{pmatrix}. \tag{77}$$

With the help of these solutions, one can form a complete basis of plane wave states

$$\phi_{p,n}^a(x, t) = \frac{1}{(2\pi\hbar)^{3/2}} u_n^a(p) e^{\frac{i}{\hbar}(p_k x_k - E^a(p)t)}, \tag{78}$$

and write the general solution of Maxwell's equations as a linear combination, with an amplitude $A^a(p)$, in the form

$$G_n(x, t) = \sum_a G_n^a(x, t) = \sum_a \int d^3 p \, A^a(p) \phi_{p,n}^a(x, t). \tag{79}$$

This decomposition will allow us to identify the part of the solution that has to be eliminated due to the original additional.

In the state with the eigenvalue $E^0 = 0$ the helicity is 0, i.e., the spin is perpendicular to the direction of propagation $\mathbf{n}(p)$. The associated eigenvector $u^0(p)$ describes a movement parallel to $\mathbf{n}(p)$, i.e., a longitudinal wave. In the states with $E^\pm = \pm c p$ the helicity is $\pm 1$, that is, the spin is parallel or antiparallel to $\mathbf{n}(p)$. Both associated eigenstates are perpendicular to $\mathbf{n}(p)$, and also perpendicular to each other [the relation $\left(u^a(p), u^b(p)\right) = \delta_{ab}$ holds true for a inner product defined by $(a, b) = \sum_k a_k^* b_k$]. The movement therefore takes place in a plane perpendicular to $\mathbf{n}(p)$ and describes a transverse wave.

The transition from the original particle picture to the present quantum picture led in a certain way to a simplification, insofar as only three directions of the original continuum of possible directions of velocity have now survived. On the other hand, at first glance, a difficulty of intuitive interpretation arises in that the operators of spin and velocity are proportional to each other. The velocity operator therefore has the "wrong direction", so to say. However, the velocity is described by the eigenstate belonging to the operator of the velocity, and not by the operator itself. Therefore, the quantum mechanical state corresponding to the original particle state [**v** parallel to **n**$(p)$] is given by the longitudinal contribution $G_n^0(x, t)$ associated with the eigenvalue $E^0 = 0$.

To eliminate the term $G_n^0(x, t)$ from Eq. (79), it is sufficient to postulate the validity of the relation

$$\frac{\partial G_n}{\partial x_n} = 0. \tag{80}$$

The longitudinal state is not compatible with this requirement due to the relation $\mathbf{n}(p) = \mathbf{u}^0(p)$, while the transversal states automatically fulfil this condition. After separating the real and imaginary parts, (80) leads to the transversality conditions

$$\frac{\partial \mathbf{E}}{\partial \mathbf{r}} = 0, \qquad \frac{\partial \mathbf{B}}{\partial \mathbf{r}} = 0 \tag{81}$$

for the real-valued physical fields **E** and **B**. The conditions (81) play a special role; as is well-known, it is sufficient to ensure their validity at a single (initial) time. This special role is also visible in the formalism of second quantization.

The absence of the longitudinal mode for massless particles can also be shown using Lorentz invariance; the same is true for the parallelism between the velocity operator and the spin operator [26]. It is satisfying that the HLLK provides an alternative and completely independent explanation for these facts.

## 8 Maxwell's equations, classical or quantum

We have already identified the system of Eqs. (72),(73),(81) with Maxwell's equations, but we may ask ourselves whether this identification is the only possible one within the framework of our theory.

In Sect. 5, we introduced a density whose physical meaning and dimension was not specified. The formalism of HLLK yields the same field equations for all these densities. However, the physical meaning and dimension of the associated fields is determined by that of the densities; the dimension of the density coincides with the dimension of the square of the fields. Equations (72),(73) imply the conservation law

$$\frac{1}{2c}\frac{\partial}{\partial t}\left(\mathbf{E}^2 + \mathbf{B}^2\right) + \frac{\partial}{\partial \mathbf{r}}\left(\mathbf{B} \times \mathbf{E}\right) = 0. \tag{82}$$

It follows that the density $\mathbf{E}^2 + \mathbf{B}^2$ integrated over whole space is a conserved quantity. If we interpret the fields **E**, **B** as electrodynamical fields, then this density is an energy density. In order to derive Maxwell's equations, one must therefore take a non-probabilistic quantity, namely an energy density, as the original density.

This choice is consistent with the interpretation of Maxwell's equations as classical equations, where *classical* means the same as non-probabilistic. The HLLK thus provides the basis for a "second quantization" in the sense of the quantization of a classical field. As will be discussed in the next section, the process of HLLK can be thought of as a generalized quantization. If this terminology is accepted, the field quantization becomes a *second* quantization in the literal sense.

Regardless of the fact that there is no true probability density in Maxwell's equations, the 3-component quantity $\psi$ introduced in Sect. 7.2 (or a generalization of it) is called the "wave function of the photon" by some authors [27]; this terminology has been criticized by other authors [28]. The question is not of paramount importance in view of the fact that numerous quantum-like structures can be found in Maxwell's equations, especially in the formulation of Eq. (68). These structures also became visible in the course of our derivation.

As already mentioned, the interpretation of the density, which is first introduced in the phase space, is completely arbitrary in the HLLK, and has no influence on the form of the final field equations in the configuration space. However, if a probability interpretation is assigned to the density, then one has to fulfil the additional condition that the integral of the density over the whole space must be equal to 1 (a similar operation exists also in the case of a particle density). For example, using the formalism introduced in I, II, we can also derive a classical Schrödinger equation where the square of the absolute value of the wave function has the meaning of a particle density. Conversely, instead of the classical Maxwell equations derived here, we can also derive a "quantum mechanical" equation, for a three-component complex quantity $\phi$, with associated fields $\mathbf{b}$ and $\mathbf{e}$. These fields satisfy Eqs. (72),(73),(81) and, in addition, must satisfy the probabilistic normalization condition

$$\int d^3x \left(\mathbf{b}^2 + \mathbf{e}^2\right) = 1, \tag{83}$$

with the integration extending over all space. The relation between this quantum-mechanical Maxwell wave function and the above classical wave function was found by Good [25,28]. He also showed that the expectation values of Maxwell's theory have the correct quantum mechanical form when using this wave function.

These questions are relevant for the problem of second quantization, the introduction of an occupation-number formalism for a first-quantized theory, or the equivalent problem of quantizing a classical field [29]. The additional complication that the relevant dynamical variables are not the fields but the potentials must also be taken into account [30]. We do not want to discuss these questions, which are outside the scope of this paper, but mention that the HLLK provides a basis for both types of second quantization. We hope to return to these questions in a later paper, after the derivation of the Dirac equation.

## 9 Discussion

In I–IV, Schrödinger's equation (and the entire formalism of QT as regards single particles) was derived with the help of the HLLK. In the present work, the complete set of Maxwell's equations was derived using the same method. It was thus possible to derive the two differential equations that are probably the most important in the field of fundamental physics with the help of this method.

These two equations are quite different in their physical meaning. We know from an enormous number of observations that the predictions of Schrödinger's equation are probabilistic in nature whereas the predictions of Maxwell's equations are deterministic, in the sense that they can be tested with the help of forces in single experiments. Schrödinger's equation thus belongs to QT, while the electromagnetic field is regarded as the prototype of a classical field par excellence; at least as long as one excludes experiments with very low intensity from consideration.

The HLLK was originally constructed to derive QT within the framework of a probabilistic world view. The fact that the same method can now be used to derive a deterministic (fundamental) system of classical field equations requires a reassessment of the significance of this method. One can define the HLLK as a generalized quantization method that starts from a continuum of particle trajectories, performs a projection onto the configuration space, and introduces a new natural constant in the course of this projection. In the case of QT, this new natural constant is $\hbar$, while in the case of Maxwell's equations it is the speed of light $c$.

This double success leads to the speculative question whether possibly *all* fundamental fields of physics can be defined on the basis of such a generalized quantization defined by the HLLK. It leads, in short, to the question of whether all fundamental fields are quantum fields in this sense. This would also mean that all fundamental fields of physics can be traced back to an associated particle system. With regard to the dichotomy mentioned at the beginning of this paper, it would mean that the particle concept is more fundamental than the field concept.

The natural constants generated in the course of the projection to the configuration space, $\hbar$ and $c$, both have the physical meaning of a realistic limitation or—which is the same thing—the elimination of an unrealistic idealization. Let us explain what we mean by that. Classical particle physics contains the unrealistic assumption that measurements with infinitely high accuracy are possible. The HLLK quantization that leads to Schrödinger theory eliminates this unrealistic assumption and creates an accuracy limit $\hbar$. As regards the massless case, let us consider the structure of Galilean space-time. Here we find the unrealistic assumption that infinitely large velocities are possible. The HLLK quantization eliminates this unrealistic assumption by generating the Maxwell field—and with it a modified, relativistic space-time structure characterized by a maximum velocity $c$.

We find here a new ordering principle for the relation between different physical theories: Theory $B$ is *better* than theory $A$ if an inadmissible idealization of theory $A$ no longer occurs in theory $B$. In the present work, we have derived—following a path opened by Holland [9]—Minkowski space from Galilean space with the help of this ordering principle. We can now transfer classical mechanics (for particles with non-zero mass) into Minkowski space according to the well-known method. This relativistic mechanics still suffers from the inadmissible idealization that measurements with infinitely high accuracy are possible. The elimination of this inadmissible idealization, with the help of the HLLK quantization, leads to the Dirac equation, as will be shown in the next paper in this series.

**Funding**  Open access funding provided by Johannes Kepler University Linz.

**Availability of data and materials**  Data are contained within the article.

**Declarations**

**Conflict of interest**  There is no conflict of interest.

## Appendix A Proof of hermiticity of Hamilton operator

Here we show that the Hamiltonian (49) is Hermitian with respect to the inner product (48) defined for $2\pi$-periodic functions. The proof is carried out as usual by partial integration, whereby the angle variables $\alpha_k$ must be taken into account in addition to the position variables $x_k$; to separate the variables, we write $d\sigma = d^3x\, dS$, $dS = d^3\alpha\, \sin\alpha_1$. First, starting from the definition

$$\left(\psi, \hat{H}\phi\right) = -\frac{s_0^2}{d_0} \int_{\mathbb{R}^3} d^3x \int_{SO(3)} dS\, \psi^*(x,\alpha) A_{lk}(\alpha) \frac{\partial}{\partial\alpha_l} \frac{\partial}{\partial x_k} \phi(x,\alpha), \tag{A1}$$

we perform the partial integrations with respect to the variables $x_k$ and obtain

$$\left(\psi, \hat{H}\phi\right) = \frac{s_0^2}{d_0} \int_{\mathbb{R}^3} d^3x \int_{SO(3)} dS\, \left[\frac{\partial}{\partial x_k}\psi^*(x,\alpha)\right] A_{lk}(\alpha) \frac{\partial}{\partial\alpha_l} \phi(x,\alpha). \tag{A2}$$

After carrying out the three remaining partial integrations with respect to the variables $\alpha_k$, taking into account the $2\pi$-periodicity of the wave functions, we obtain the relation

$$\left(\psi, \hat{H}\phi\right) = \int_{\mathbb{R}^3} d^3x \int_{SO(3)} dS\, \left(\hat{H}\psi\right)^* \phi + \Delta, \tag{A3}$$

where $\Delta$ results from the angular dependence of the matrix $A$ and is given by

$$\Delta = -\frac{s_0^2}{d_0} \int_{\mathbb{R}^3} d^3x \int_{SO(3)} dS\, \left(\frac{\partial}{\partial x_k}\psi\right)^* \left[\frac{\partial A_{lk}}{\partial\alpha_l} + \frac{\cos\alpha_1}{\sin\alpha_1} A_{1k}\right]. \tag{A4}$$

A direct calculation shows that all three components of the square bracket in Eq. (A4) vanish. Therefore, $\Delta = 0$ and $\left(\psi, \hat{H}\phi\right) = \left(\hat{H}\psi, \phi\right)$.

## References

1. Einstein, A.: Autobiographical notes. In: Schilpp, P.A. (ed.) Albert Einstein: Philosopher-Scientist, p. 2. Harper and Row, New York (1949)

2. Feldmeier, A.: Theoretical Fluid Dynamics. Springer, Cham (2019)
3. Klein, U.: From probabilistic mechanics to quantum theory. Quantum Stud. Math. Found. **7**, 77–98 (2020)
4. Caratheodory, C.: Calculus of Variations and Partial Differential Equations of the First Order, Part I. Holden-Day, Inc, San Francisco (1965)
5. Ballentine, L.E.: The statistical interpretation of quantum mechanics. Rev. Mod. Phys. **42**, 358–381 (1970)
6. Klein, U.: From Koopman–von Neumann theory to quantum theory. Quantum Stud. Math. Found. **5**, 219–227 (2018)
7. Klein, U.: A reconstruction of quantum theory for nonspinning particles. arXiv:2202.13356
8. Klein, U.: A reconstruction of quantum theory for spinning particles. arXiv:2202.13364
9. Holland, P.: Hydrodynamic construction of the electrodynamic field. Proc. R. Soc. A **461**, 3659–3679 (2005)
10. Bondar, D.I., Cabrera, R., Lompay, R.R., Ivanov, M.Y., Rabitz, H.A.: Operational dynamic modeling transcending quantum and classical mechanics. Phys. Rev. Lett. **109**, 190403 (2012)
11. Arsiwalla, X.D., Chester, D., Kauffman, L.H.: On the operator origins of classical and quantum wave functions. Quantum Stud. Math. Found. **11**, 193–215 (2024)
12. Chester, D., Arsiwalla, X.D., Kauffman, D., Planat, L.H., Irwin, K.: Quantization of a new canonical, covariant, and symplectic Hamiltonian density. Symmetry **16**(3), 316 (2024)
13. Barandes, J.A.: The stochastic-quantum correspondence. arXiv:2302.10778
14. Barandes, J.A.: The stochastic-quantum theorem. arXiv:2309.03085
15. Wetterich, C.: Probabilistic cellular automaton for quantum particle in a potential. arXiv:2211.17034
16. Ohanian, H.C.: What is spin? Am. J. Phys. **54**, 500–505 (1986)
17. Goldstein, H., Poole, C., Safko, J.: Classical Mechanics. Addison-Wesley Press Inc., San Francisco (2001)
18. Holland, P.R.: The Quantum Theory of Motion. Cambridge University Press, Cambridge (1995)
19. Sudarshan, E.C.G., Mukunda, N.: Classical Dynamics: A Modern Perspective. Wiley, New York (1974)
20. Madelung, E.: Quantentheorie in hydrodynamischer Form. Z. Phys. **40**, 322–326 (1926)
21. Whittaker, E.T.: A Treatise on the Analytical Dynamics of Particles and Rigid Bodies. Cambridge University Press, London (1989)
22. Raymer, M.G., Smith, B.J.: The Maxwell wave function of the photon. Proc. SPIE **5866**, 293 (2005)
23. Field, J.H.: Relationship of quantum mechanics to classical electrodynamics and classical relativistic mechanics. Eur. J. Phys. **25**, 385–397 (2004)
24. Wigner, E.P.: Group Theory and Its Application to the Quantum Mechanics of Atomic Spectra. Academic Press, New York (1959)
25. Good, R.H.: Particle aspect of the electromagnetic field equations. J. Phys. Rev. **105**, 1914–1919 (1957)
26. Wigner, E.P.: Relativistic invariance and quantum phenomena. Rev. Mod. Phys. **28**, 255–268 (1957)
27. Bialynicki-Birula, I.: Photon wave function. In: Wolf, E. (ed.) Progress in Optics, vol. XXXVI, pp. 1–46. Elsevier, Amsterdam (1996)
28. Kiessling, M.K., Tahvildar-Zadeh, A.S.: On the quantum-mechanics of a single photon. J. Math. Phys. **59**, 112302 (2018)
29. Sebens, C.T.: Electromagnetism as quantum physics. Found. Phys. **49**, 365–389 (2019)
30. Akhiezer, A.I., Berestetskii, V.B.: Quantum Electrodynamics. Interscience Publishers, New York (1965)